# Deposition-Dependent Coverage and Performance of Phosphonic Acid Interface Modifiers in Halide Perovskite Optoelectronics

Hannah Contreras,[1] Aidan O'Brien,[1,†] Margherita Taddei,[1] Yangwei Shi,[1,2] Fangyuan Jiang,[1] Robert J. E. Westbrook,[1] Yadong Zhang,[3] Rajiv Giridharagopal,[1] Paul A. Lee,[4] Stephen Barlow,[3] Seth R. Marder,[3,5] Neal R. Armstrong,[4*] David S. Ginger[1*]

1. Department of Chemistry, University of Washington, Seattle, WA 98195, USA

2. Molecular Engineering & Sciences Institute, University of Washington, Seattle, WA 98195, USA

3. Renewable and Sustainable Energy Institute, University of Colorado-Boulder, Boulder, CO 80309, USA

4. Department of Chemistry and Biochemistry, University of Arizona, Tucson, AZ 85721, USA

5. Department of Chemical and Biological Engineering and Department of Chemistry, University of Colorado-Boulder, Boulder, CO 80309, USA

† Deceased February 18th, 2025.

*Corresponding authors: nra@arizona.edu, dginger@uw.edu

KEYWORDS: Perovskite solar cells, self-assembled monolayer, interfaces, contact engineering, phosphonic acid modification

*Abstract*

In this work, we study the effect of various deposition methods for phosphonic acid interface modifiers commonly pursued as self-assembled monolayers in high-performance metal halide perovskite photovoltaics and light-emitting diodes. We compare the deposition of (2-(3,6-diiodo-9H-carbazol-9-yl)ethyl)phosphonic acid onto indium tin oxide (ITO) bottom contacts by varying three parameters: the method of deposition, specifically spin coating or prolonged dip coating, ITO surface treatment via $HCl/FeCl_3$ etching, and use in combination with a second modifier, 1,6-hexylenediphosphonic acid. We demonstrate that varying these modification protocols can impact time-resolved photoluminescence carrier lifetimes and quasi-Fermi level splitting of perovskite films deposited onto the phosphonic-acid-modified ITO. Ultraviolet photoelectron spectroscopy shows an increase in effective work function after phosphonic acid modification and clear evidence for photoemission from carbazole functional groups at the ITO surface. We use X-ray photoelectron spectroscopy to probe differences in phosphonic acid coverage on the metal oxide contact and show that perovskite samples grown on ITO with the highest phosphonic acid coverage exhibit the longest carrier lifetimes. Finally, we establish that device performance follows these same trends. These results indicate that the reactivity, heterogeneity, and composition of the bottom contact help to control recombination rates and therefore power conversion efficiencies. ITO etching, prolonged deposition times for phosphonic acids via dip coating, and the use of a secondary, more hydrophilic bis-phosphonic acid, all contribute to improvements in surface coverage, carrier lifetime, and device efficiency. These improvements each have a positive impact, and we achieve the best results when all three strategies are implemented.

*TOC:*

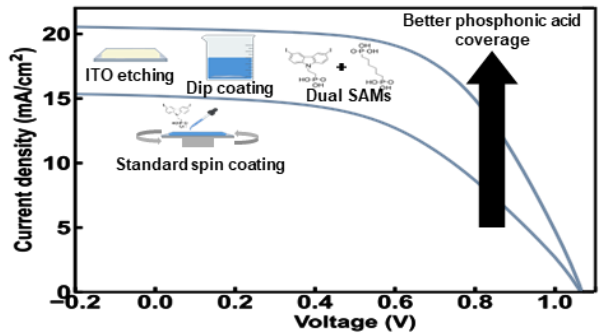

**Introduction**

Metal halide perovskites have emerged as next-generation semiconductors with applications ranging from light-emitting diodes,[1,2] to photovoltaics,[3] and even quantum light.[4] In energy harvesting applications, perovskites demonstrate power conversion efficiencies (PCEs) competitive with silicon,[5] tunable optoelectronic properties, and potential for low-cost, scalable fabrication.[6–9] Recent improvements in the efficiency and operational stability of perovskite light-emitting diodes,[10,11] solar cells,[12,13] and organic optoelectronic devices[14–16] are, in part, due to the incorporation of phosphonic acid interface modifiers into the device architecture.[2,17–23] These phosphonic acids are commonly used to tailor hole-extracting (or hole-injecting) indium tin oxide (ITO) bottom contacts in solar cells and perovskite LEDs.[2,24–26] Phosphonic acid modifiers show strong binding affinities with ITO and other metal oxide electrodes, forming layers – often referred to as self-assembled monolayers (SAMs) – that mitigate the extreme chemical and electrical heterogeneity of ITO,[27–30] passivate interfacial defects to reduce charge recombination,[31] tailor the ITO work function to better match the valence band energy of the active layer,[32] and control interfacial energies and wetting between the active layer and the contact, thereby improving overall device efficiency.[27–29,33–39] We postulate that heterogeneity within the surface of ITO poses challenges to achieving full passivation of these contacts utilizing phosphonic acids, where thermodynamic and kinetic barriers may exist in the reactivity toward modifiers such as phosphonic acids. Furthermore, regions of high electrical activity, if not fully passivated may be sites that promote surface recombination, loss of perovskite solar cell efficiency, and in some cases may be sites of electrochemical activity that helps to control long term stability.[40]

Often, studies report device performance and other optoelectronic parameters without reference to the exact deposition methods or their effects on the surface coverage and orientation of the phosphonic acid interface modifiers.[41] Indeed, in many cases there are assertions that the phosphonic acids form self-assembled monolayers; this terminology popularized for other systems such as thiols on gold where the ability of the thiol to move on the surface and desorb and reabsorb allows for highly -ordered monolayers (e.g., of alkanethiols on gold)[42] may not be applicable to the binding of phosphonic acids on many metal oxides. Also, the details of the washing process must be considered carefully with regards to drawing conclusions as to whether the phosphonic acid truly forms a monolayer.

Other recent studies suggest that variation of phosphonic acid modifier deposition techniques, especially those which lead to increased phosphonic acid coverages at the buried interface, can lead to significant performance gains in both organic and perovskite solar cells.[43–49] Intriguing recent work by Aalbers et al., using a variety of phosphonic acid interface modifiers on ITO contacts, has shown that a simple six -carbon alkyl diphosphonic acid outperforms examples of the carbazole-phosphonic acid modifiers more typically used in perovskite solar cell platforms in terms of mitigating shallow defects at the ITO/perovskite interface and improving photoluminescence (PL), time-resolved photoluminescence (trPL) and quasi-Fermi level splitting (QFLS) in perovskite active layers.[50] From these and related studies it is clear that more detailed considerations need to be given to the nature of such interface modification schemes, including pathways to enhance modifier coverage, the role of bottom contact activation prior to the modification process, and the effects of combining interface modifiers in mitigating surface wettability issues, interface-driven nonradiative recombination and ultimately enhancing performance in perovskite solar cell platforms.

ITO contacts tend to be quite heterogeneous on sub-micron to nanometer length scales, in terms of their topologies (sub-grain structure and proportions of the low index faces of the bixbyite lattice which are exposed at the surface), their electrical and electrochemical properties, and chemical reactivities toward modifiers such as phosphonic acids.[27,27–29,33–39]

Herein, we investigate the impact of various ITO activation methods and phosphonic acid deposition techniques on the surface coverage of the carbazole phosphonic acid [2-(3,6-diiodo-9H-carbazol-9-yl)ethyl]phosphonic acid (I-2PACz), a model carbazole-PA surface modifier, either alone or in combination with the 6-carbon PA, 1,6-hexylenediphosphonic acid (6dPA). We explore the effects of these ITO contact modifiers on wide bandgap ($E_g$ = 1.7 eV) perovskite solar cell devices. We use ultraviolet photoelectron spectroscopy (UPS) and X-ray photoelectron spectroscopy (XPS) to characterize the resulting surfaces, and we find that (1) the activation of the ITO surface using a short, aggressive chemical etch,[33] (2) the prolonged substrate exposure to the phosphonic acid solution through a 12-hour liquid-phase deposition,[28,51] and (3) the addition of the bifunctional 6dPA molecule as a second modifier, all contribute systematically to increased phosphonic acid modifier coverage on ITO. We show that perovskites grown on films with improved phosphonic acid modifier coverage exhibit decreased nonradiative charge carrier recombination and increased quasi-Fermi level splitting under illumination. As a result, solar cells made using all three treatments together show the highest power conversion efficiencies.

**Results and Discussion**

*Studies of surface modification protocols for phosphonic acids on ITO*

We first explore the effects of I-2PACz as a model surface modifier. I-2PACz is a member of a group of carbazole-phosphonic acid modifiers that are widely used on ITO contacts in low bandgap p-i-n perovskite solar cells,[20,24,25] and more recently, LEDs.[2] We also use 6dPA as a reference phosphonic acid because of its success as an additive increasing surface energies in combination with other modifiers,[19] and more recently as an effective passivation layer to lower surface recombination rates on ITO contacts.[50] In **Figure 1a and 1b** we show the structures of the two phosphonic acids. Using these two molecules, we prepare three compositions of interface modification layers according to **Figure 1c** consisting of (1) I-2PACz, (2) 6dPA, and (3) a film made by sequentially depositing first I-2PACz and then 6dPA. Within these three compositions, we further compare two different phosphonic acid deposition conditions: first, spin coating of the phosphonic acid, a popular approach for deposition on ITO substrates,[41] and second, a 12-hour dip coating deposition that resembles earlier systematic studies of phosphonic acid layer deposition.[28,51] In addition, these modification protocols are carried out for two different surface cleaning/activation protocols of ITO prior to treatment with the phosphonic acids: a conventional oxygen plasma treatment,[52] and a 10 second wet chemical etching step.[33]

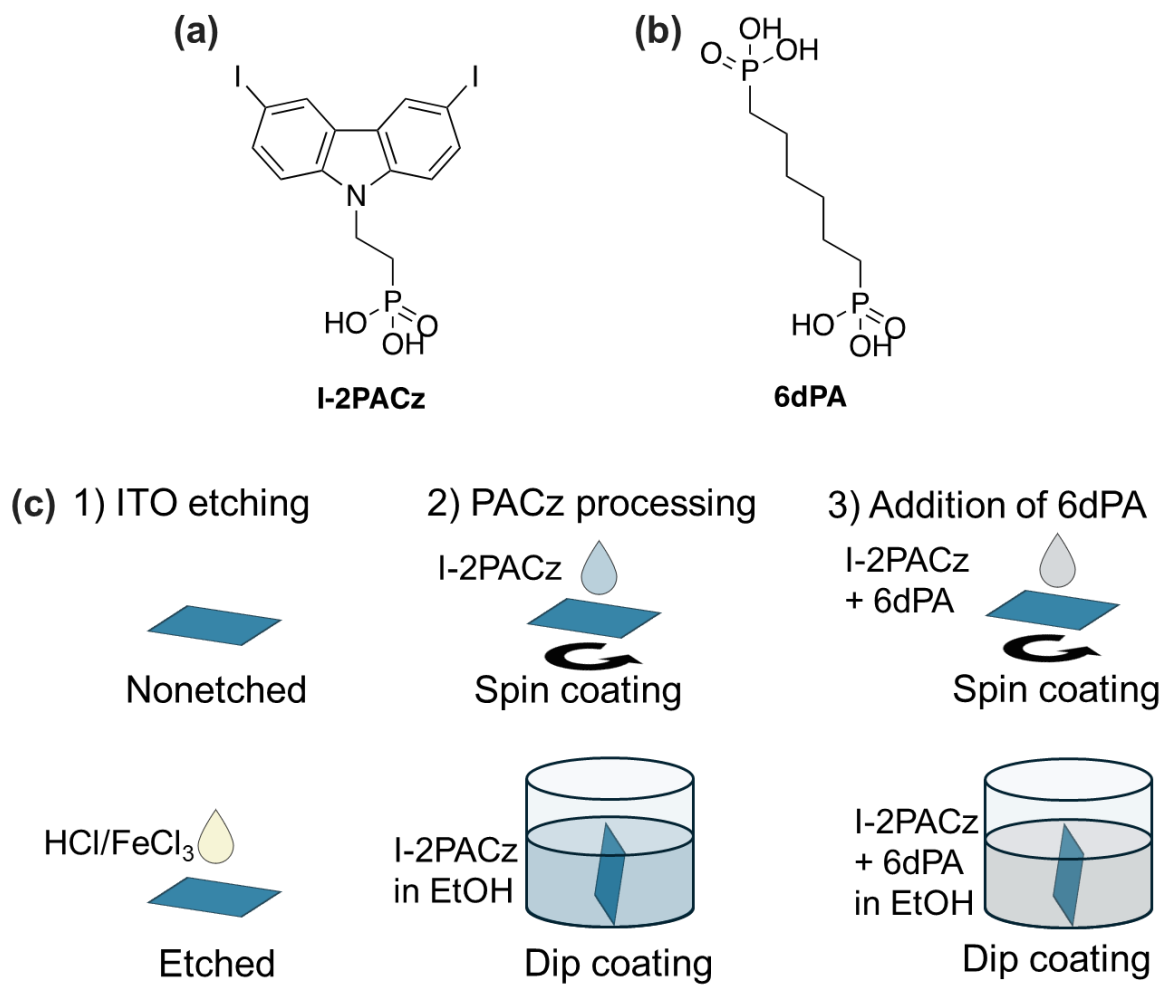

*Figure 1.* Structures of the phosphonic acid modifiers (a) I-2PACz and (b) 6dPA and (c) the surface activation and modification processes used in this study.

We first compare the effects of the different phosphonic acid compositions, deposition methods, and ITO treatments on QFLS and the trPL lifetime of perovskite deposited from solution on top of ITO modified in a variety of ways. We use $FA_{0.83}Cs_{0.17}Pb(I_{0.75}Br_{0.25})_3$ "Cs17Br25" perovskite as a model wide bandgap (1.7 eV) active layer due to the interest in wider-gap top cells in p-i-n architectures and the use of metal oxides, including ITO, in perovskite/silicon tandem devices.[26,48,53]

**Figure 2a** shows the QFLS of each ITO/phosphonic acid/perovskite half-stack deduced from the steady state PL quantum yields measured under AM1.5 equivalent intensity. All QFLS values are reported in **Table S1.** The QFLS improves 2.5% from 1.16 ± 0.01 eV (± 95% confidence interval) of the bare ITO/perovskite to 1.19 ± 0.02 eV with the inclusion of the nonetched spin coated I-2PACz layer. The QFLS increases 5.2% when etching ITO, then 6.9% when using dip-coating and 7.8% when adding 6dPA. The QFLS in these half-stacks improves with each additional surface treatment. The highest QFLS of 1.25 ± 0.02 eV is reached when all surface treatments of etching, dip-coating and additional 6dPA are applied. The addition of 6dPA contributes to greater QFLS as well. A larger QFLS, determined optically through the photoluminescence quantum yield (PLQY), suggests that devices made with these same surface modifications could achieve higher external potential, or open circuit voltage ($V_{OC}$), if they were contacted optimally.

We hypothesize that these different surface treatments may alter QFLS because of changes in nonradiative recombination at the modified ITO/perovskite interface, as has been proposed previously for phosphonic acid modified-interfaces.[50] Recently, Aalbers et al. found that 6-carbon-bis-phosphonic acid provides better surface passivation compared to other more commonly used carbazole phosphonic acids.[50]

To study the kinetics of nonradiative recombination at the surface, we use trPL.[54,55] In **Figure 2b and 2c**, we show the trPL decays for each surface treatment. We see that etched ITO, modified with I-2PACz deposited by prolonged dip coating, and dual I-2PACz / 6dPA samples show higher carrier lifetimes compared to their counterparts fabricated using nonetched ITO, spin coated I-2PACz, and single phosphonic acid layers. PL decays for films fabricated on nonetched ITO (**Figure 2b**) are consistently faster than those of films fabricated on etched ITO (**Figure 2c**). Faster PL decays indicate lower carrier lifetimes (extracted through stretched exponential fitting), more PL quenching, and more nonradiative recombination pathways at the ITO/phosphonic acid/perovskite interface.[55] Longer lifetimes in combination with increased PLQYs indicate better passivated interfaces.

Overall trPL trends are consistent with the results from the PLQY measurements, showing an over 55x improvement in trPL lifetime as interface modification is optimized. Spin coating I-2PACz on nonetched ITO results in the fastest decay with the most nonradiative recombination and lowest carrier lifetime of 3.8 ns ± 2.6 ns. ITO etching prior to modification reduces nonradiative recombination and increases the lifetime of the perovskite/spin coated I-2PACz sample to 18.9 ns ± 5.8 ns. Prolonged dip coating of etched ITO substrates in I-2PACz instead of spin coating I-2PACz improves the ITO/perovskite interface further and carrier lifetime increases to 92.3 ns ± 8.1 ns. Prolonged dip coating of etched ITO in I-2PACz solution and subsequently a 6dPA solution achieves the best average lifetime of 210 ns ± 28 ns.

These increases in photoluminescence lifetime with changes to the ITO surface treatment indicate that the resulting perovskite films grown on the different surfaces exhibit progressively lower densities of energetically shallow defects and fewer nonradiative recombination events.[50] We postulate that this type of interface modification minimizes direct contact between the perovskite and regions of the ITO responsible for faster surface recombination, (electrical hot spots that have been observed in conducting tip AFM characterization),[29,33] but also possibly because the perovskite grown from those surfaces itself has fewer defects.[56] As a control, we also prepared perovskite films on substrates with 6dPA only.

While trPL lifetimes are rather short (16.2 ± 3.8 ns) for perovskites grown on spin coated 6dPA films on nonetched ITO, those grown on nonetched ITO dip coated in 6dPA are actually better (longer) than those grown on dip coated I-2PACz / 6dPA. Films grown on etched ITO dip coated 6dPA have lifetimes similar to (close to being within experimental uncertainty of) but still shorter than the lifetimes of films grown on their dip coated I-2PACz / 6dPA counterparts. This comparison suggests that, at least in terms of improving PL lifetime, the carbazole functionality in these phosphonic acid interface modifiers is not essential, and that differences in surface coating/film properties arising from the etching and dip coating are more important. We next turn to surface characterization via UPS/XPS to better understand both alterations in surface coverage and energetics as modification protocols are altered.

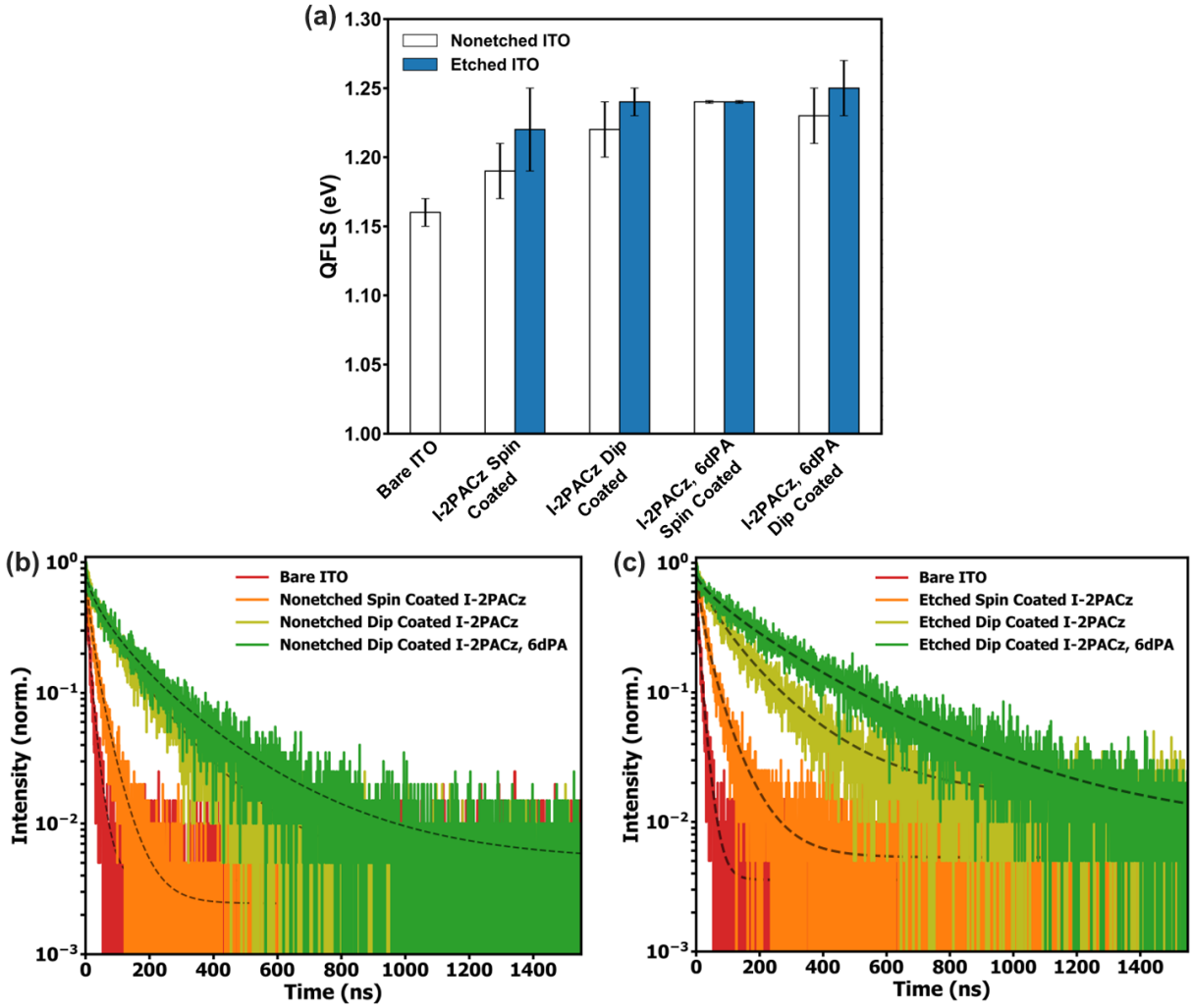

***Figure 2.*** *(a) QFLS for half-stack ITO/phosphonic acid/perovskite samples with the phosphonic acid layers deposited using different approaches on nonetched (white) and etched (blue) ITO substrates. Error bars represent the 95% confidence interval across at least six films. TrPL decays for the same (b) nonetched and (c) etched ITO/phosphonic acid/perovskite samples.*

***Characterization of clean, etched and phosphonic acid coverage of ITO using UPS and XPS***

We acquire UPS data to estimate effective work function and valence band maxima changes as we added phosphonic acid layers to both nonetched and etched ITO contacts. Comparison with the trPL and QFLS data above show that work function changes alone cannot explain the improvements in those parameters. **Table 1** summarizes our UPS measurements of the average effective work functions of the different phosphonic acid modified ITO surfaces.[27] **Figures S2-S8** in Supporting Information show the relevant spectra and the x,y point-to-point variability in the low kinetic energy (LKE) edge distributions that we routinely see for these phosphonic acid modified ITO contacts. We expect phosphonic acids to modify the work function of ITO due to the vacuum level shift induced by their interface dipole.[16,32,47] However, the magnitude of this shift depends on the net dipole moment of the phosphonic acid, the molecular density (surface coverage) and can be especially sensitive to the position (tilt angle) of the functional group(s) closest to the modifier/vacuum interface.[42,57,58]

**Table 1.** *UPS work function data depending on phosphonic acid modifier deposition method (± 95% confidence interval). Two values are reported for the pure 6dPA samples because of variability in their LKE edges.*

| Work Function | Nonetched (eV) | Etched (eV) |
|---|---|---|
| ITO control | 4.08 ± 0.11 | 4.03 ± 0.01 |
| ITO / I-2PACz spin | 4.63 ± 0.15 | 4.77 ± 0.11 |
| ITO / I-2PACz prolonged dip coat | 5.10 ± 0.20 | 5.21 ± 0.13 |
| ITO / I-2PACz / 6dPA spin | 4.91 ± 0.09 | 5.01 ± 0.15 |
| ITO / I-2PACz / 6dPA prolonged dip coat | 4.64 ± 0.02 | 4.25 ± 0.19 |
| ITO / 6dPA spin | 4.41 ± 0.09, 4.54 ± 0.10 | - |
| ITO / 6dPA prolonged dip coat | - | 4.40 ± 0.20, 4.48 ± 0.13 |

All modifiers increase the effective work function of both nonetched and ITO surfaces by up to 1.2 eV. For modification with just I-2PACz the work function increase appears to correlate with the higher surface coverages suggested by the XPS data discussed below. The work function of spin coated I-2PACz increases from 4.63 eV on nonetched ITO to 4.77 eV on etched ITO. When we deposit I-2PACz via prolonged dip coating, the work function further increases to 5.10 eV on nonetched ITO, and up to 5.21 eV on etched ITO. Importantly, these increases in work function indicate increased modifier coverage results from both ITO etching and prolonged dip coating. Because the work function shift is proportional both to the number density of phosphonic acids and their relative tilt angle on the surface, these results suggest that etching and prolonged dip coating result in overall higher surface coverage.[27] In confirmation of that hypothesis, photoemission from the carbazole π-orbitals is clearly resolved (BE = ca. 6.3 eV) and highest intensity for I-2PACz modification of the etched ITO surface (**Figures S4 and S5**) as expected if the carbazole groups dominate photoemission within the sampling depth (ca. 1 nm) of the He I photoemission experiment. This increase in surface coverage in turn is consistent with the trPL lifetime and QFLS above.

Prior to the deposition of 6dPA, the effective work function of the I-2PACz modified ITO contacts increases with coverage of pure I-2PACz, consistent with changes in surface coverage of each modifier leading to alterations of the net local dipoles that control the position of the LKE edge in the photoemission spectra. The effective work function of pure 6dPA deposited onto etched ITO by prolonged dip coating is 4.5 eV. For the 6dPA phosphonic acid coating we note inflection points in the LKE edge that suggest non-uniform "layers," exhibiting two distinct effective work functions depending upon region of ITO examined. This type of inflection in LKE edges has been documented previously for "patchy" thin film coatings.[59]

In addition, when we deposit I-2PACz and 6dPA sequentially by prolonged dip coating, the resulting work function on etched ITO samples is 4.3 eV. The observation of a lower work function for the mixed composition layer than for either pure I-2PACz or pure 6dPA suggests that the measured work functions are not a simple average of the two work functions corresponding to the two molecular dipoles and that complex rearrangements of these mixed phosphonic acids are likely. Because these complex surfaces still provide good trPL lifetimes and improved QFLS, it is clear that work function alone cannot be determining these properties.

Using XPS data we qualitatively gauge the thickness of the I-2PACz modifying layer by monitoring the attentuation of the In $3d_{5/2}$ peak (area) after each modification step (**Figure S10** and **Table S6**), assuming an inelastic mean free path, IMFP = 3 nm, for the In 3d photoelectrons, and normal (0°) takeoff angles. Although we observe significant variability in the absolute intensities, the following clear trends emerge which correlate with the differences in QFLS and trPL shown above: a) for nonetched ITO samples, increasing the deposition time from spin coating to prolonged dip coating correlates a near doubling of the effective thickness. We observe the same increase in I-2PACz coverage (even for spin coated modifiers) when comparing the reactivity of the etched ITO versus the control. This change demonstrates that more chemically reactive oxide surfaces vastly outperform control surfaces at shorter phosphonic acid modifier exposure times; b) the addition of 6dPA to the ITO/I-2PACz surfaces does not significantly increase the effective coverage, even though XPS lineshapes indicate a clear difference in composition (see **Figure 3** below); and c) the acid-etched ITO samples again achieve the largest overall effective thickness of the modifier layer.

We next use XPS to characterize the resulting phosphonic acid layers in detail. We focus on the peak area ratios to delineate substrate coverage. **Figures 3a-d** show how the O 1s, P 2p, N 1s, and I 3d XPS spectra progress for each phosphonic acid surface treatment. We focus here on the acid-etched ITO samples because of their superior PL performance. **Figures S9-S14** and **Tables S7-S10** provide additional XPS spectra for both etched and nonetched ITO substrates and full peak area ratios.

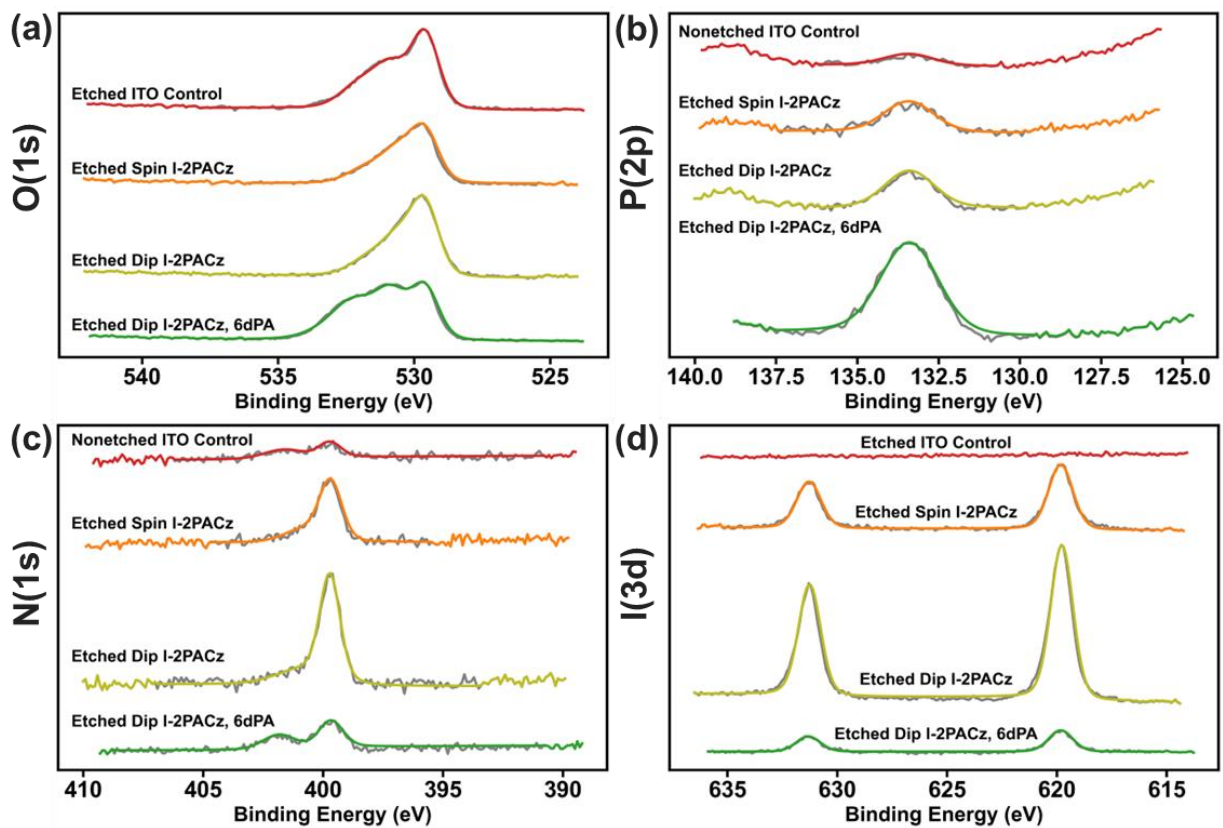

*Figure 3.* XPS spectra of control and phosphonic acid modified acid-etched ITO: (a) O 1s, (b) P 2p, (c) N 1s, and (d) I 3d. The gray background represents the raw data and the fitted data are overlaid in color.

The O 1s spectrum in **Figure 3a** shows three distinct features in the O 1s spectra. The lowest binding energy peak (529.6 eV) on the unmodified, etched ITO surface arises from oxygen in the bixbyite indium oxide lattice[33,35] and shows substantial asymmetry to the high binding energy side of the peak owing to energy loss processes (scattering of photoelectrons) typically seen in highly doped metal oxide semiconductors.[33] The subsequent (symmetric) higher binding energy features extending from 530.5 to 531.6 eV arise from a distribution of hydroxyl groups on the surface of the oxide and oxygen arising from trace adsorbed $H_2O$, respectively. The heterogeneity of types and concentrations of such surface -OH groups on indium oxide and ITO surfaces is well established,[33,38,60] we can assume a broad distribution of such sites, and in their reactivities toward reactions with phosphonic acids.[27,28,35] I-2PACz introduces O 1s photoemission arising from -HO-P-O-In- near-surface species, corresponding to monodentate, bi-dentate and possibly tri-dentate bonding of the phosphonic acid to the oxide and overshadowing the high binding energy O 1s features from the ITO.[35] The relative concentration of -OH-like O 1s photoemission features is higher for spin coated I-2PACz layers versus those created by prolonged dip coating, consistent with the hypothesis that the prolonged dip coating process provides more complete phosphonic acid attachment to the oxide surface.

When we add 6dPA to this more complete dip coated I-PACz layer, the O 1s envelope increases in complexity because of the photoemission from the oxygen in exposed phosphonic acid groups at the surface, and possibly even phosphonic acid networks in regions with high local 6dPA concentrations, consistent with the unusual trends in work function we showed above on samples with extended 6dPA coating and clear evidence for multilayered regions on these surfaces. Significantly, the O/In relative atomic ratios (peak area ratios corrected for relative sensitivity factors detailed in **Table S7**) steadily increase from the etched bare ITO control to spin coated I-2PACz to prolonged dip coated I-2PACz samples, with the highest ratios achieved for dual I-2PACz / 6dPA mixed layers, indicating greater phosphonic acid coverage with each surface treatment, in agreement with the trPL results.

**Figure 3b** shows the shape of the P 2p peak at 133.4 eV is unchanged for all phosphonic acid modifier deposition techniques, indicating that all deposited modifiers have a similar array of binding modes for phosphorous in these layers.[35] Modifier coverage increases in prolonged dip coated and dual modifier samples as we observed for the O 1s series, determined by P/In peak area ratios. We summarize these results in **Table S6**. For the I-2PACz/6dPA mixed layers the P 2p peak intensity is much larger, due to the upward facing phosphonic acid head group component, and formation of heterogeneous PA multilayers.

**Figure 3c** shows that the main N 1s peak appears at 399.7 eV and increases in area from the spin coated to prolonged dip coated samples, further evidence that more molecules are deposited by the prolonged dip coating method. The N signal is dampened by the addition of 6dPA, obscuring the trend in these samples. A small higher binding energy-shoulder peak might be attributable to formation of a carbazolium cation, from partial oxidation of the carbazole layer because of the reactive oxide surface, but further studies would be required to confirm this reactivity. **Figure 3d** shows the I $3d_{5/2}$ peak at 619.8 eV intensity and displays the same peak area trend as the N 1s data. I-2PACz layers created from prolonged dip coating produce higher I $3d_{5/2}$ intensities and the inclusion of 6dPA in the modifying layer diminishes these intensities.

The I/In, N/In, and P/In peak area ratios increase for the pure I-2PACz cases when deposited by prolonged dip coating instead of spin coating. The I/In and N/In ratios decrease with prolonged dip coating after the inclusion of 6dPA, suggesting that 6dPA may form a multilayer. Atomic ratios from angle resolved XPS (ARXPS), especially the I/In ratios at the two takeoff angles (**Table S10**), also support the formation of a 6dPA multilayer or disordered I-2PACz layer of nonuniform orientation. If the deposited I-2PACz layer orients primarily with the phosphonic acid group down, bound to the ITO surface and the iodide pointing upwards for engagement with the perovskite, the atomic P ratio should decrease and the atomic I ratio should increase in the high angle (60°) scans, differentiating them from that of the 0° controls. However, apart from the nonetched I-2PACz sample, the atomic P ratio increases and the atomic I ratio decreased with high angle scans, indicating that the I-2PACz layer has a somewhat disordered orientation in both spin coating and prolonged dip coating even before the addition of 6dPA, suggesting that, even in these samples, it is likely that some areas of multilayer coverage exist. **Tables S7-S10** detail both 0° and 60° ARXPS data.

*Exposure to PbI$_2$:DMF/DMSO solution and chemisorption of PbI$_x$*

The role played by the unmodified regions of the ITO surface, and regions where phosphonic acid groups are exposed to solution, is unresolved. In preliminary experiments, we use XPS to observe the uptake of PbI$_x$ intermediates from DMF/DMSO containing PbI$_2$ on both bare ITO and I-2PACz modified ITO. These experiments suggest that clean ITO surfaces and even ITO/I-2PACz surfaces modified with the highest coverages of phosphonic acid, have accessible sites for uptake of Pb. We posit that adsorbed Pb is another indication of the chemical heterogeneity of the ITO and phosphonic acid-modified ITO surfaces and a possible indicator of potential recombination sites even on modified oxide surfaces.

Adsorption of heavy metals like Pb on ITO or related transparent conductive oxides has not been reported, but the adsorption on Al$_2$O$_3$ surfaces, with similar Lewis and Bronsted acid-base chemistry, and comparable speciation of surface hydroxyls (-OH) is well recognized.[33,61] On hydrated Al$_2$O$_3$ surfaces adsorption of Pb occurs through interactions with deprotonated surface -OH groups and we posit that similar adsorption processes occur with clean and acid-etched ITO surfaces, and that any available -OH sites on even the phosphonic acid modified surfaces can act as an adsorption site for Pb.

We expose control ITO surfaces and ITO surfaces modified with I-2PACz and I-2PACz / 6dPA to a 10$^{-3}$ M PbI$_2$-DMF:DMSO solution, well below the concentrations used in creating perovskite active layers to ensure these interactions came from solutions that were not close to saturation.

**Figures S16 and S17** show the Pb 4f, S 2p and I 3d spectra for these surfaces. We see the uptake of Pb at low coverages for bare ITO surfaces, I-2PACz/ITO surfaces and more than double these coverages when 6dPA is introduced as an additional modifier. No additional sulfur is seen on any of these surfaces, and I/Pb ratios for adsorption on clean, etched and nonetched ITO sample is less than 2.0, suggesting adsorption of a DMSO-free, sub-stoichiometric $PbI_x$. For I-2PACz modified ITO surfaces, the apparent coverage of Pb is comparable to that seen on clean and acid-etched ITO which suggests only a small change in adsorption capacity despite the presence of the phosphonic acid modifier. This observation is consistent with a constant percentage of the ITO surface available for Pb uptake even at high I-2PACz coverages. For surfaces additionally modified with 6dPA there is a substantial increase (2×) in overall Pb coverage and for photoemission data taken at 60° takeoff angles an even larger increase (4×) is seen, consistent with adsorbed Pb concentrated at the upper surface of the modifying layer, adsorbed through phosphonic acid-Pb interactions. This indicates that different phosphonic acid modifiers impact the deposition of the perovskite overlayer differently. **Tables S9 and S10** provide the full data for Pb adsorption on bare and phosphonic acid modified ITO samples.

*Characterization of photovoltaic responses for modified ITO/Cs17Br25 device platforms*

The sum of the photoelectron and photoluminescence experiments suggest that the different deposition methods result in different phosphonic acid coverages, and that these coverages directly affect the photocarrier dynamics of the perovskite semiconductor layers that grow on top of them. To further verify the impact of the phosphonic acid coverage on the optoelectronic performance of the perovskite, we fabricate working solar cell stacks of the structure ITO/phosphonic acid/Cs17Br25/C60/BCP/Ag. **Figure 4a** shows a schematic of the device architecture, and **Figures**

**4b-4f** show the resulting device current–voltage (*J*–*V*) curves, $V_{OC}$, short circuit current ($J_{SC}$), fill factor (FF), and PCE distributions.

In **Figure 4c** we see that the $V_{OC}$ improves slightly for devices built on etched ITO. Notably, despite increases in the trPL lifetime and the quasi-Fermi level splitting of the partial stacks, we observe relatively little improvement in $V_{OC}$ in our devices. We interpret this result to indicate that our devices are limited more by film quality than by surface recombination dynamics, an interpretation consistent with the $J_{SC}$ and FF data below.

The performance effect of the modifier deposition method is more apparent when considering the $J_{SC}$. **Figure 4d** shows that etching alone improves the $J_{SC}$ of the nonetched spin coated I-2PACz control from $17.29 \pm 0.59$ mA/cm$^2$ to $18.03 \pm 0.65$ mA/cm$^2$. Etching and prolonged dip coating improves this further to $19.73 \pm 0.74$ mA/cm$^2$. Etching, prolonged dip coating, and dual I-2PACz / 6dPA deposition result in the highest average $J_{SC}$ of $20.45 \pm 1.22$ mA/cm$^2$, a more than 15% increase in $J_{SC}$.

**Figure 4e** shows that, while phosphonic acid type and deposition method have a slight effect on the FF, etching improves the FF significantly and consistently. On average, I-2PACz based devices have 31% higher FF when deposited on etched ITO. Etched, dip coated I-2PACz / 6dPA devices obtain the highest average FF, $0.54 \pm 0.07$. Given that fill factor changes most with etching and is less sensitive to the exact type of phosphonic acid, we propose that the FF changes simply result from better ITO activation which results in higher sunt and lower series resistance, consistent with the trPL and XPS results.

Finally, **Figure 4f** summarizes the resulting PCE values. ITO etching, prolonged dip coating, and a dual modifier layer each separately improve PCE. After we applied all three processing enhancements, the devices achieved a top PCE of 14.35%, a 57% increase from the best nonetched spin coated I-2PACz control. Although the trPL and QFLS behavior of perovskite on pure 6dPA films is among the best samples in this study, perovskite devices grown on pure 6dPA alone do not result in good devices, as we show in the Supporting Information along with all other device statistics, confirming that while high quality perovskite can be grown on insulating substrates, the work function and electronic transport properties of the interface nevertheless still matter for device performance in these higher bandgap perovskite active layers. We recognize that while we identify clear, systematic methods to improve the buried interface, these devices still fall short of the state-of-the-art perovskite solar cell efficiencies.[20,47]

As an additional control, we also explore the generality of modified deposition conditions with more commonly used, higher-performing modifiers such as (2-(3,6-dimethoxy-9H-carbazol-9-yl)ethyl)phosphonic acid and (4-(3,6-dimethyl-9H-carbazol-9-yl)butyl)phosphonic acid (MeO-2PACz and Me-4PACz, respectively). With these more common modifiers, we increase performance from 9.75 ± 2.13% with spin coating Me-4PACz on etched ITO to 18.40 ± 1.01% with dip coating Me-4PACz and 6dPA on etched ITO without any other modifications to enhance wettability (**Table S17 and S18**). This result underscores that an enormous amount of variability in the literature likely exists due to variations in ITO and surface preparation.

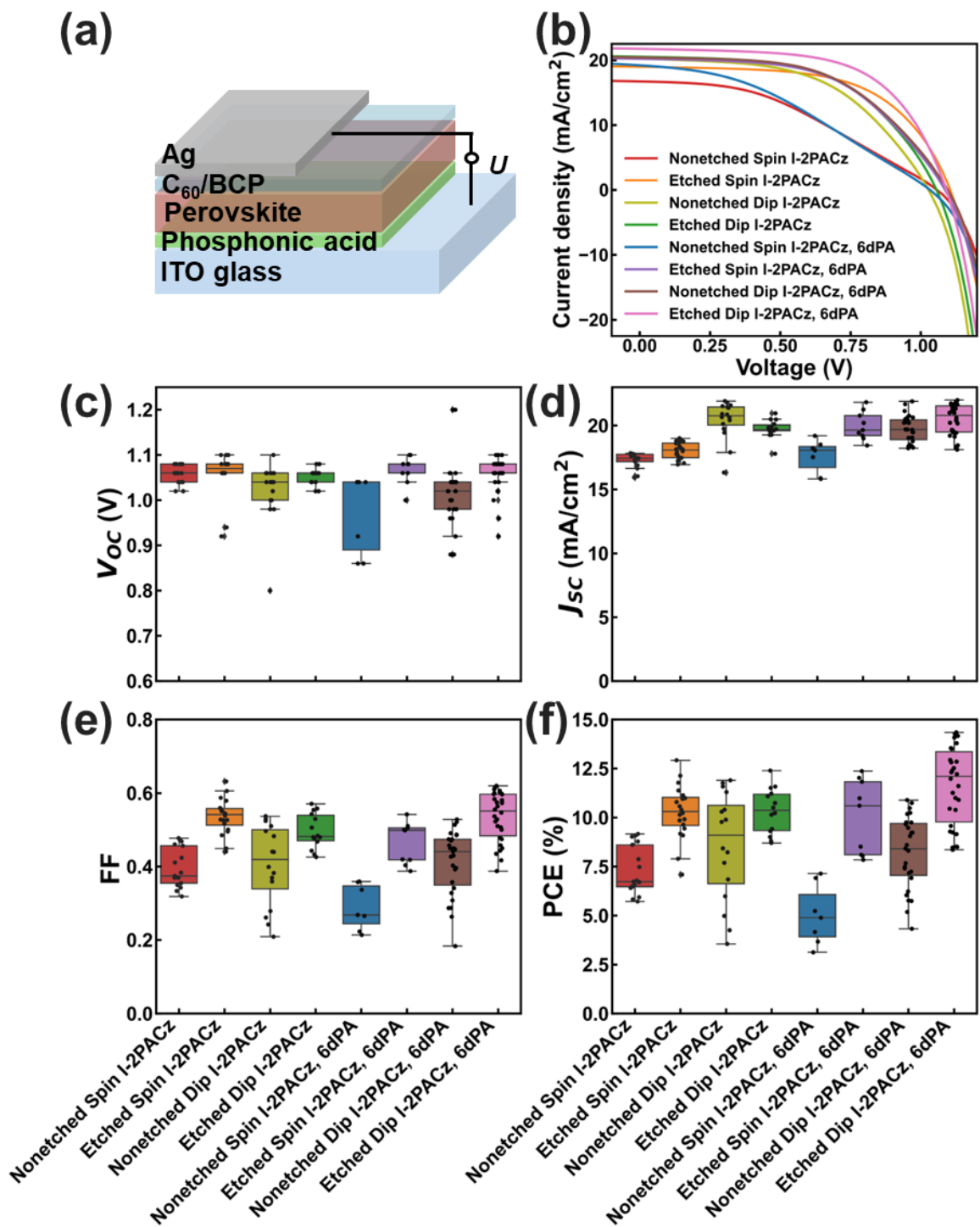

*Figure 4.* (a) The device stack architecture and statistical performance parameters (b) J–V curves, (c) $V_{OC}$, (d) FF, (e) $J_{SC}$, and (f) PCE for each of the phosphonic acid deposition methods: nonetched vs. etched, spin coated vs prolonged dip coated, and single I-2PACz vs. dual I-2PACz / 6dPA

*modifiers. The line in the center of each box indicates the average and the box edges signify one quartile from the average.*

**Conclusion**

We establish a direct correlation between the deposition of phosphonic acid modifiers, their coverage on ITO, and the performance of solar cell devices. Highly doped ITO has become the standard bottom contact for a wide range of optoelectronic devices, but it is clear that the heterogeneity in its reactivity with phosphonic acids and related modifiers needs to be understood and mitigated for creation of perovskite solar cells at scale. Our results suggest that ITO chemical etching, prolonged solution coating, and the use of a dual phosphonic acid layer all improve phosphonic acid coverage on ITO, consistent with the enhanced work function and peak area ratio progressions. This improved phosphonic acid coverage in turn leads to reduced nonradiative decay in perovskite films grown on higher quality phosphonic acid layers, and consequently to improved photovoltaic device performance. The improved coverage likely modulates the electronic interactions at the ITO/perovskite interface, while also homogenizing the intrinsic heterogeneity of ITO surfaces. The use of mixed phosphonic acids improves coverage and interfacial wetting. This understanding of tailoring the buried interface is beneficial at both the laboratory and commercial scale, as it provides a template path for reliably optimizing deposition for these molecules while bypassing their historical wettability (coverage-based) issues. While our study focuses on the surface coverage and electronic properties of the interface, we also find evidence that the phosphonic acid layers can affect pre-adsorption of Pb to the surface. We speculate that this is a further indication of the availability of adsorption sites on the oxide even after phosphonic acid modification which needs to be accounted for in subsequent efforts to minimize the population of potential surface recombination sites and for control of nucleation and growth of optimized perovskite active layers.

**Acknowledgments**

This work is dedicated to the memory of Aidan O'Brien, whose intelligence, dedication, and insight were invaluable to its completion. This work began with funding under Office of Naval Research (Award # N00014-20-1-2587) but was primarily supported by Hannah Contreras's work under a National Science Foundation Graduate Research Fellowship under Grant No. DGE-2140004. Any opinion, findings, and conclusions or recommendations expressed in this material are those of the author(s) and do not necessarily reflect the views of the National Science Foundation. Atomic force microscopy and wide-bandgap perovskite semiconductor growth was supported by the U.S. Department of Energy, Office of Basic Energy Sciences, Division of Materials Sciences and Engineering under Award DE-SC0013957. We also acknowledge the Rabinovitch Endowed Fund of the Department of Chemistry at the University of Washington for materials and supplies. Resynthesis of I-2PACz for this work was performed at CU Boulder and was supported by the U.S. DOE EERE under SETO (Award No. DE-EE0010502). We acknowledge support by the State of Washington through the University of Washington Clean Energy Institute and the Washington Research Foundation. The X-ray and ultraviolet photoelectron spectra in this work were collected at the Laboratory for Electron Spectroscopy and Surface Analysis (LESSA) in the Department of Chemistry and Biochemistry at the University of Arizona using a Kratos Axis 165 Ultra DLD Hybrid Ultrahigh Vacuum Photoelectron Spectrometer. The instrument was purchased with funding from the National Science Foundation and supported by the Center for Interface Science: Solar-Electric Materials (CIS: SEM), an Energy Frontier Research Center funded by the U.S. Department of Energy, Office of Basic Energy Sciences under award number DESC0001084.